\documentclass[aps,preprintnumbers,showkeys,showpacs]{revtex4}
\usepackage{epsfig}
\usepackage{psfrag}
\usepackage{amsfonts}
\usepackage{graphicx}
\usepackage{dcolumn}
\usepackage{bm}
\usepackage{setspace}
\usepackage{multirow}

\begin{document}
\title{Revisit cosmic ray propagation by using $^{1}$H, $^{2}$H, $^{3}$He and $^{4}$He}

\author{Juan Wu}
\email{wu@cug.edu.cn}
\author{Huan Chen}
\affiliation{School of Mathematics and Physics, China University of Geosciences, Wuhan 430074,
China}

\begin{abstract}
The secondary-to-primary ratios are unique tools to investigate cosmic ray propagation mechanisms. In this work, we use the latest data of deuteron-to-helium~4 ratio and helium~3-to-helium~4 ratio measured by PAMELA combined with other Z$\leq$2 primary fluxes measured by PAMELA and Voyager-1, to constrain the cosmic ray acceleration and propagation models. The analysis is performed by interfacing statistical tools with the GALPROP propagation package. To better fit both the modulated and unmodulated low energy cosmic ray data, we find that a time-, charge- and rigidity-dependent solar modulation model is better than the force-field approximation. Among all the studied cosmic ray propagation models, the diffusion-reacceleration-convection model is strongly supported by the derived Bayesian evidence. The robustness of the estimated diffusion slope $\delta$ is cross-checked by another low-mass secondary-to-primary ratio, i.e. the antiproton-to-proton ratio. It is shown that the diffusion-reacceleration-convection model can reconcile well with the high energy antiproton-to-proton ratio. This indicates that the estimated value of $\delta$ is reliable. The well constraint $\delta$ from the ``best'' model is found to be close to 1/3, inferring a Kolmogorov-type interstellar magnetic turbulence.
\end{abstract}

%\pacs{98.70Sa}
\maketitle

\section{Introduction}\label{sec:intro}
Understanding cosmic ray (CR) acceleration and propagation mechanisms is an attractive topic over the last century. The standard CR theory reveals that Galactic CRs are accelerated via Fermi acceleration in supernova shocks, consequently diffuse on turbulence in Galactic magnetic field and meanwhile possibly experience reacceleration and convection~\cite{Strong2007}. Though this paradigm is generally accepted, the details of the involved physical processes remain unclear. For instance, the spectral index of the diffusion coefficient presented in recent literatures differs from 0.2 to 0.6~\cite{Johannesson2016, Korsmeier2016, Yuan2017}, which could indicate a big difference on the properties of Galactic magnetic turbulence. Recently, benefiting from the launches of various space missions such as PAMELA~\cite{Picozza2007}, AMS-02~\cite{Aguilar2013} and DAMPE~\cite{Chang2017}, CR physics has entered a data-driven epoch. These new instruments circling around Earth for a long duration and suffering no residual atmosphere interference, eliminate the statistical and systematic uncertainties of the measurements. Besides, a spacecraft Voyager-1 launched at 1977, has reached the boundary of Heliosphere since 2012 and can directly observe interstellar CRs. By using the unprecedented precise data measured by these experiments, we are allowed to investigate the acceleration and propagation phenomena elaborately and reliably.

In order to probe CR propagation models, the most commonly used
secondary-to-primary ratio in previous theoretical studies was the
boron-to-carbon (B/C) ratio. The diffusion-reacceleration (DR)
model was usually proposed which could naturally explain the peak
around 1~GeV in the B/C ratio (e.g. see~\cite{Moskalenko2002,
Strong2004, Trotta2011, Johannesson2016}). However, a discrepancy
between the antiproton ($\bar{\text{p}}$) data and the DR model
prediction (see \cite{Trotta2011} and references therein) can't be
ignored. It was suggested in \cite{Johannesson2016} that the
$Z\leq2$ CR species might suffer different propagation properties
with respect to the heavier elements. Consequently, an evident
problem is that using only the B/C ratio may bias our evaluation
on CR transport models. Some studies argued that the
``$\bar{\text{p}}$ discrepancy'' could be caused by the inaccurate
$\bar{\text{p}}$ production cross-sections and new cross-sections
were proposed~\cite{diMauro2014, Kachelriess2015, Winkler2017,
Donato2017}. Incorporating different hadronic models can lead to
an uncertainty on $\bar{\text{p}}$ production achieving
50\%~\cite{Lin2017}. Since the $\bar{\text{p}}$ problem is far
from certain, alternatively, it is very important and necessary to
employ other low-mass secondaries to investigate CR propagation
mechanisms.

The deuterium-to-helium~4 ($^2$H/$^4$He) ratio and the helium~3-to-helium~4 ($^3$He/$^4$He) ratio give us this opportunity, which were shown to be a powerful tool to constrain CR propagation models~\cite{Coste2012, Tomassetti2012, Picot2015}. They have been measured by several experiments, such as the balloon borne experiments IMAX92~\cite{deNolfo2000, Menn2000}, BESS~\cite{Wang2002, Kim2013}, BESS-Polar II~\cite{Picot2017} and the space instruments AMS-01~\cite{Aguilar2011} and PAMELA~\cite{Adriani2016}. Unlike the balloon experiments, the PAMELA measurements did not have systematic uncertainties caused by the atmospheric background. And it provided data with better precision and wider energy range than AMS-01. Therefore, in this paper, we use the $^2$H/$^4$He and $^3$He/$^4$He ratios measured by PAMELA as well as other $Z\leq2$ primary spectra to constrain CR acceleration and propagation models.

\section{Data} \label{sec:data}

Combining the secondary-to-primary ratios and the primary spectra allows us to explore both propagation and acceleration processes. The $^2$H/$^4$He ratio from 0.1~GeV/n to 1.4~GeV/n and the $^3$He/$^4$He ratio from 0.1~GeV/n to 1.1~GeV/n published by PAMELA are used in our fittings. These two types of ratios at higher energies were only measured by IMAX92, but with very limited accuracies, which are not used here. For the primary spectra, PAMELA has measured the proton (p) flux from 400~MeV to 1.2~TeV and the helium (He) flux from 100~MeV/n to 600~GeV/n~\cite{Adriani2011}. These data will also be included in this work. We note that AMS-02 has released the p and He fluxes with better statistics~\cite{Aguilar2015a, Aguilar2015b}. But these data were taken from May of 2011 to November of 2013. A reversal of Heliospheric magnetic field (HMF) polarity from $A<0$ to $A>0$ occurred between October of 2012 and June of 2013. During this period, the HMF varied dramatically and the solar effects could be very complicated. In order to facilitate the solar modulation calculation, the AMS-02 p and He data are not utilized in this analysis.

Apart from the PAMELA data, we also use the CR spectra measured by
Voyager-1~\cite{Stone1977}. It is for the first time that
scientists have observed interstellar CR spectra. Elemental
intensities for the $Z\leq2$ particles provided by Voyager-1
include the p flux from 3~MeV/n to 346~MeV/n and the He flux from
3~MeV/n to 661~MeV/n~\cite{Stone2013, Cummings2016}. The prominent
roles of the Voyager-1 data can be stressed as follows. First, CRs
measured by Voyager-1 suffer no burden from Solar wind, which can
solely constrain CR acceleration and propagation models without
considering the influence of solar modulation. Second, by
comparing the modulated data to the unmodulated data, the solar
modulation effect can be probed straightforwardly.

\section{Solar modulation} \label{sec:solar}

Solar modulation is one of the main obstacles for us to extract message from CR data observed on Earth. In literatures, a most frequently used model for solar modulation was the force-field approximation~\cite{Gleeson1968}. This model depends on one free parameter, the so-called Fisk potential $\Phi$. Though this model has been widely used due to its easiness in usage, it does not give a reliable description on the heliospheric transport. Recently, some studies treated this phenomenon in a more reasonable way~\cite{Bisschoff2016, Cholis2016, Ghelfi2016, Corti2016, Boschini2017}. For example, in~\cite{Boschini2017}, a solar propagation code HelMod~\cite{Bobik2012, Bobik2016} was incorporated with the CR transport code GALPROP~\cite{Moskalenko1998, Strong1998} to calculate the CR spectra. This approach allowed a physical calculation of solar modulation. However, the inter-calibrating step between GALPROP and HelMod was complicated and time-consumed. Meanwhile, HelMod performs a Monte Carlo integration for each CR species to calculate the modulated spectrum. The Monte Carlo simulation for deuteron has not been available yet.

Another heliospheric model was proposed in~\cite{Cholis2016}, based on several well measured time-dependent solar observables. Considering a CR particle with charge $q$, rigidity $R$ and velocity $\beta=\upsilon/c$, this model constructs the potential $\Phi$ as:
\begin{eqnarray}
\centering
\Phi \left(R,t\right)&=&\phi_{0}\left(\frac{\left|B_{\text{tot}}\right|}{4\;\text{nT}}\right)+\phi_{1}H\left(-qA\left(t\right)\right)
\left(\frac{\left|B_{\text{tot}}\right|}{4\;\text{nT}}\right)\nonumber\\
&\times&\left(\frac{1+\left(R/R_{0}\right)^{2}}{\beta\left(R/R_{0}\right)^{3}}\right)\left(\frac{\alpha\left(t\right)}{\pi/2}\right)^{4},
\label{eq:CholisSolar}
\end{eqnarray}
where $A$ and $\left|B_{tot}\right|$ are the polarity and
magnitude of the HMF measured at Earth, $\alpha\left(t\right)$ is
the tilt angle of the heliospheric current sheet (HCS), $R_{0}$ is
a reference rigidity, $\phi_{0}$ and $\phi_{1}$ are two
normalization factors. It introduces a Heaviside step function
$H\left(-qA\left(t\right)\right)$ which equals to zero or unity,
depending on the sign of $qA\left(t\right)$. The second term of
Equation~\ref{eq:CholisSolar} represents a drift movement along
the HCS, which is the main process suffered by particles with
$qA\left(t\right)<0$. For $qA\left(t\right)>0$, CRs travel rather
directly from the polar regions of the heliopause to Earth, and
$H\left(-qA\left(t\right)\right)=0$. This model, referred as CM
model in the following paragraphs, is physically motivated and can
be interfaced with GALPROP as easy as the force-field
approximation.

In this work, we attempt to interface the CM model with GALPROP to perform a simultaneous scan on all the involved parameters. By comparing the results evaluated separately from the CM model and from the force-field model, we can understand better the uncertainties caused by solar modulation. The PAMELA measurements employed in this work were derived from data collected between July of 2006 and December of 2008. As observed, $\left|B_{\text{tot}}\right|$ and $\alpha$ vary with time~\cite{ACE2018,WSO2018}. To determine each CR spectrum over the whole period, we calculate the modulated spectrum for each individual six-month duration and then average them. We set $R_{0}$ to 0.5~GV~\cite{Cholis2016}. It was found in~\cite{Cholis2017} that the estimated values of $\phi_{1}$ varies significantly by considering different CR propagation models. The estimation of $\phi_{0}$ may also be relevant with the CR source and propagation parameters chosen in the analysis. Therefore, the factors $\phi_{0}$ and $\phi_{1}$ are both allowed to be free in our fittings.

\section{Methodology}
\subsection{Model parameter description}\label{sec:paras}

To solve the CR propagation equation, we use the version r2766
\footnote{https://sourceforge.net/projects/galprop/} of GALPROP.
We focus our study on the transport and source parameters
characterising different processes described as follows. Diffusion
is a consequence of CRs interacting with magnetohydrodynamic
turbulence. The diffusion coefficient $D_{xx}$ is estimated to be
a power law in rigidity $\rho$~\cite{Ptuskin2006, Strong2007}, and
is usually assumed as:
\begin{equation}
\centering
D_{xx}=D_{0}\beta^{\eta} \left(
\frac{\rho}{\rho_{0}}\right)^{\delta}, \label{eq:diff_coeffieient}
\end{equation}
where $D_{0}$ is the normalization of diffusion coefficient at a reference rigidity $\rho_{0}$, $\beta$ is the particle velocity, $\eta$ is a low energy dependence factor which could possibly caused by the turbulence dissipation effect~\cite{Ptuskin2006} and $\delta$ is the diffusion spectral index. The free parameters linked to diffusion include $D_{0}$, $\delta$ and $\eta$.

Diffusion happens not only in position space, but may also in momentum space, which results in reacceleration of CR particles. The associated diffusion coefficient in momentum space $D_{pp}$ is correlated to $D_{xx}$, and is expressed as~\cite{Seo1994}:
\begin{equation}
\centering
D_{pp}=\frac{4v_{A}^{2}p^{2}}{3\delta\left(4-\delta^{2}\right)\left(4-\delta\right)D_{xx}},
\end{equation}
where $v_{A}$ is the velocity of fluctuations in the hydrodynamical plasma, called the \textit{Alfv$\acute{e}$n velocity}. $v_{A}$ is the main free parameter related to reacceleration.

Besides, CR particles may transport in bulk away from Galactic plane by the convection process. The convection velocity $V_{c}(z)$ is usually assumed to vary linearly with the distance from Galactic plane:
\begin{equation}
\centering
V_{c}\left(z\right)=V\left(0\right)+\frac{\mathrm{d}V}{\mathrm{d}z}z,
\end{equation}
here $V(0)$ and $\mathrm{d}V/\mathrm{d}z$ are the main free parameters.

According to the diffusive shock acceleration theory, the injected density $q$ for CR species $i$ is assumed to follow a rigidity power law. The general form is given as:
\begin{equation}
\centering
q_{i} = N_{ i} f \left(R,z\right) \rho^{-\nu},
\end{equation}
where $f\left(R,z\right)$ is the spatial distribution of sources and $N_{i}$ is the normalization abundance for CR species $i$. Some studies suggested a broken power law of injection spectrum~\cite{Trotta2011, Cummings2016}, which employed different injection indices $\nu_{1}$ and $\nu_{2}$ above and below a reference rigidity $\rho_{br}$. Meanwhile, heavier elements may have different indices with protons. We assume that $\nu_{1}$ and $\nu_{2}$ are the injection indices for protons, and $\nu_{1\text{He}}$ and $\nu_{2\text{He}}$ are those for helium nuclei. Their differences are set as:
\begin{equation}
\centering
\nu_{1\text{He}}=\nu_{1}+\Delta_{\text{He}};\quad\nu_{2\text{He}}=\nu_{2}+\Delta_{\text{He}}.
\label{eq:injindex}
\end{equation}
The parameters $\nu_{1}$, $\nu_{2}$ and $\Delta_{\text{He}}$ are allowed to vary freely. In GALPROP, the source abundance of protons $N_{\text{p}}$ is normalized based on the propagated proton spectrum at 100~GeV, and $N_{i}$ for other species are scaled by their abundances relative to that of protons. In our previous study~\cite{Wu2012}, we found that for different models, $N_{p}$ varied slightly. Thus we set $N_{\text{p}}=4.69\times10^{-9}$\,cm$^{-2}$\,sr$^{-1}$\,s$^{-1}$\,MeV$^{-1}$ to fit the PAMELA proton data. The helium abundance relative to proton is used as a free parameter, i.e. $X_{\text{He}}$. As suggested in~\cite{Picot2015}, the production cross-sections of $^2$H and $^3$He provided in~\cite{Coste2012} are adopted in this work to give better description of $^2$H and $^3$He production.

To summarize, our free parameters include the propagation parameters: $D_{0}$, $\delta$, $\eta$, $v_{A}$, $V(0)$, $\mathrm{d}V/\mathrm{d}z$; the source parameters: $\nu_{1}$, $\nu_{2}$, $\Delta_{\text{He}}$, $X_{\text{He}}$; and the solar modulation parameters: $\phi_{0}$ and $\phi_{1}$ (for the CM model), or $\Phi$ (for the force-field model). A degeneracy exists between $D_{0}$ and the halo size of Galaxy $z_{h}$. It can only be broken by unstable species.  In this work, $z_{h}$ is fixed at 4~kpc to be consistent with earlier studies~\cite{Strong2001, Trotta2011, Boschini2017} and to ease the comparison of results.

\subsection{Statistical methods}

To perform a quick understanding on how well a model reflects the
observed data, the $\chi^{2}$ ``goodness of fit'' test is used.
The minimization library MINUIT~\cite{James1975} is interfaced
with GALPROP for the analysis. This allows a direct recognization
of improper models. For those models which can generally reproduce
the data, a further Bayesian approach is used to do a comparison
on selected models. Given the data set $\mathbf{D}$ and the vector
of free parameters $\mathbf{\Theta}$ assuming a model $H$, Bayes'
theorem states that:
\begin{equation}
\centering
P \left( \mathbf{\Theta}|\mathbf{D}, H \right) =
\frac{P\left(\mathbf{D}|\mathbf{\Theta}, H\right) P
\left(\mathbf{\Theta} | H\right)}{P\left(\mathbf{D} | H\right)},
\end{equation}
where $P \left( \mathbf{\Theta}|\mathbf{D}, H \right)$ represents the posterior probability density function (p.d.f.) of the parameters, $P\left(\mathbf{D}|\mathbf{\Theta}, H\right) = L\left( \mathbf{\Theta} \right)$ is the likelihood function, $P\left(\mathbf{\Theta} | H \right)$ is the prior distribution.

The Bayesian evidence $Z =P\left(\mathbf{D} |H\right) $ is a key
ingredient to comparing alternative models. A larger value of $Z$
will be achieved by a model which depends on fewer free parameters
but fits better the data. A comparison between two hypotheses
$H_{1}$ and $H_{2}$ can be performed by comparing their respective
evidences. Assuming $Z_{2}/Z_{1}>1$, model $H_{2}$ is favored
versus model $H_{1}$, and vice versa. If the logarithm of this
ratio is larger than an empirical threshold 5.0, it represents a
strong evidence~\cite{Trotta2008}. Furthermore, the p.d.f.s of the
parameters can be derived by the Bayesian inference, which will
help us to understand the uncertainties and correlations of
parameters. To implement sufficiently fast Bayesian analysis, a
package MultiNest~\cite{Feroz2008, Feroz2009, Feroz2013} is
utilized to study CR models.

\section{Results and discussions} \label{sec:chi2}

Different types of propagation models are explored in this work: (1) the plain diffusion model (PD), (2) the plain diffusion model with an ad hoc break in the diffusion coefficient (PDbr), (3) the diffusion-reacceleration model (DR), (4) the diffusion-convection model (DC); (5) the diffusion-reacceleration-convection model (DRC). All models studied here consider a break in the injection spectrum. The data employed in the analysis only contain Z$\leq$2 data. Since $^{2}$H and $^3$He are mainly produced via interactions of $^1$H and $^4$He with interstellar medium, we let the nuclear chain start from $^{4}$He to economize the calculation consumption. A test on DRC model was carried out by setting nuclear chain begin from $^{30}$Si. It was found that the best-fit parameters only vary a few percent. This implies that ignoring the influence of heavy elements is an acceptable choice.

\subsection{A comparison between different solar modulation models} \label{sec:chi2_FF}

The CM model and the force-field model are interfaced separately with GALPROP. All the related models add a suffix ``-CM'' and a suffix ``-FF'' respectively. The results show that by incorporating the force-field approximation, none of the propagation models give a value of reduced $\chi^{2}$ close to 1. Tracing back to the contribution of each data set to the overall $\chi^{2}$, it seems that all the models are not able to simultaneously fit the modulated and unmodulated fluxes. On the other hand, for the same propagation configuration, assuming a CM scenario leads to a much smaller $\chi^{2}$ value. This suggests that compared with the force-field approximation, the rigidity-dependent CM model can describe the data better.

Since the secondary-to-primary ratios used in our fitting procedure only cover low energy data, using an incorrect solar modulation approximation may result in a bias on the evaluation of CR acceleration and propagation models, as well as the involved parameters. For example, for the force-field scenario, among all the studied models, the DR-FF model gives the smallest $\chi^{2}=2.14$. It is found that an inclusion of a convection effect does not improve the goodness of fit. But this is not the case for the CM scenario, in which the DRC-CM model gives the smallest $\chi^{2}=1.21$, as shown in Table~\ref{tab:chi2}. This model provides the best goodness of fit and prefers a strong convection with $\text{d}V/\text{d}z=30\pm5$~km\,s$^{-1}$\,kpc$^{-1}$.

\begin{table*}[!htbp] \small
\centering
\begin{tabular}{l  c  c  c  c  c  c  c }
\hline \hline
Parameter & DR-FF&PD-CM & PDbr-CM& DR-CM & DC-CM & DRC-CM \\
\hline
$D_{0} $ ($10^{28}$cm$^{2}$\,s$^{-1}$) &$8.63\pm^{0.17}_{0.16}$ & $6.14\pm0.16$ & $5.48\pm^{0.15}_{0.16}$ & $8.20\pm0.20$ & $4.3\pm0.3$ & $4.0\pm^{0.6}_{0.5}$\\
$\delta_{1}$&$0.202\pm^{0.020}_{0.019}$& $0.607\pm^{0.028}_{0.030}$ & $0.33\pm0.05$ & $0.231\pm^{0.025}_{0.023}$ & $0.80\pm^{0.05}_{0.06}$ & $0.31\pm0.04$\\
$\delta_{2}$&[$=\delta_{1}$]&[$=\delta_{1}$]& $0.84\pm0.03$ &[$=\delta_{1}$]&[$=\delta_{1}$]&[$=\delta_{1}$]\\
$\rho_{0}$ (GV)&[4]&[4]& $3.31\pm^{0.11}_{0.12}$ &[4]&[4]&[4]\\
$\eta$& $1.64\pm0.05$ & $1.27\pm0.10$ & $1.74\pm0.11$ & $1.51\pm0.06$ & $2.97\pm0.26$ & $1.52\pm0.06$\\
$v_{A}$ (km\,s$^{-1}$)&$38.4\pm^{1.6}_{1.5}$&---&---&$28.5\pm1.4$&---& $29.8\pm^{1.2}_{1.3}$\\
$V_{0}$ (km\,s$^{-1}$)&---&---&---&---&[0]&[0]\\
$\text{d}V/\text{d}z$ (km\,s$^{-1}$\,kpc$^{-1}$)&---&---&---&---&$15.6\pm^{2.3}_{2.2}$ & $30\pm5$\\
$\nu_{1}$ & $1.812\pm^{0.021}_{0.020}$ & $1.654\pm^{0.015}_{0.014}$ & $1.643\pm^{0.013}_{0.014}$ & $1.693\pm^{0.017}_{0.018}$ & $1.650\pm0.015$ & $1.741\pm^{0.079}_{0.021}$\\
$\nu_{2}$ & $2.557\pm0.018$ & $2.172\pm^{0.030}_{0.029}$ & $1.952\pm^{0.028}_{0.030}$ & $2.537\pm^{0.023}_{0.024}$ & $2.05\pm0.05$ & $2.531\pm^{0.070}_{0.029}$\\
$\rho_{br}$ (GV)& $9.6\pm^{0.5}_{0.4}$&$7.35\pm^{0.22}_{0.21}$ & $10.6\pm^{3.1}_{0.6}$ & $6.61\pm^{0.20}_{0.18}$ & $6.2\pm0.4$ & $6.81\pm^{1.21}_{0.23}$\\
$\Delta_{\text{He}}$&$-0.050\pm0.005$ &$-0.019\pm0.005$ & $-0.034\pm0.005$ & $-0.062\pm0.005$ & $-0.030\pm0.005$ & $-0.048\pm0.006$\\
$X_{\text{He}}$& $0.0696\pm0.0009$& $0.0892\pm0.0020$ & $0.1009\pm^{0.0023}_{0.0022}$ & $0.0670\pm^{0.0013}_{0.0012}$ &$0.095\pm0.004$ & $0.0705\pm^{0.0015}_{0.0014}$\\
$\phi_{0} $\multirow{2}{*}{or $\Phi$ (GV)} &\multirow{2}{*}{$0.466\pm0.006$}& $0.135\pm0.008$ & $0.152\pm^{0.009}_{0.008}$ & $0.263\pm0.012$ & $0.133\pm0.010$ & $0.218\pm^{0.013}_{0.012}$\\
$\phi_{1}$ & &$12.3\pm0.5$ & $11.2\pm^{0.5}_{0.6}$ & $8.2\pm0.7$ & $11.9\pm^{0.6}_{0.5}$ & $9.4\pm0.7$\\
$\chi^{2}/$d.o.f&2.14& 1.91 & 1.55 & 1.41 & 1.59 & 1.21\\
\hline
\end{tabular}
\caption[Best-fit parameters]{\footnotesize The best-fit parameters for DR-FF, PD-CM, PDbr-CM, DR-CM, DC-CM and DRC-CM models. The fixed parameters appear in square brackets. } \label{tab:chi2}
\end{table*}

Considering a typical value of $\phi_{1}$ around 8$\sim$12~GV
given in Table~\ref{tab:chi2}, it is clear that the rigidity
dependency of the CM model substantially influences the CR
particles with rigidities below a few~GV. At larger rigidities,
since the second term of Equation~\ref{eq:CholisSolar} only
contributes a few percent to the total value of
$\Phi\left(R\right)$, the CM scenario is almost equivalent with
the force-field model. As we knew, the index of CR spectrum at
high energies only depends on the value of $\nu_{2} +\delta_{2}$.
But it is unclear which mechanisms shape the low energy spectra.
Using the CM solar modulation model may help us to clarify this
uncertain question.

\subsection{The main results for the CM scenario}

\begin{figure}[!htb]
\centering
\includegraphics[width=8.5cm]{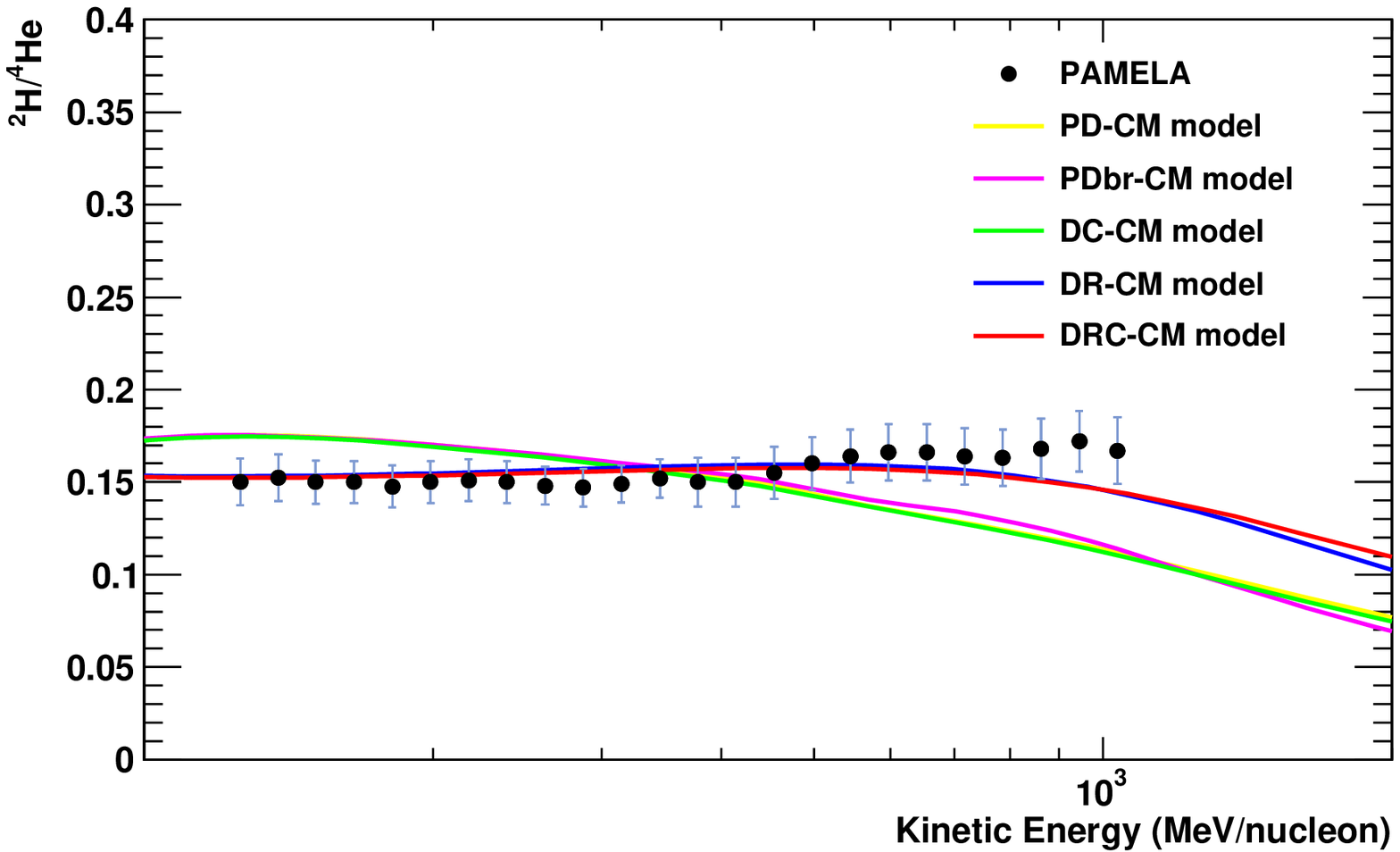}
\includegraphics[width=8.5cm]{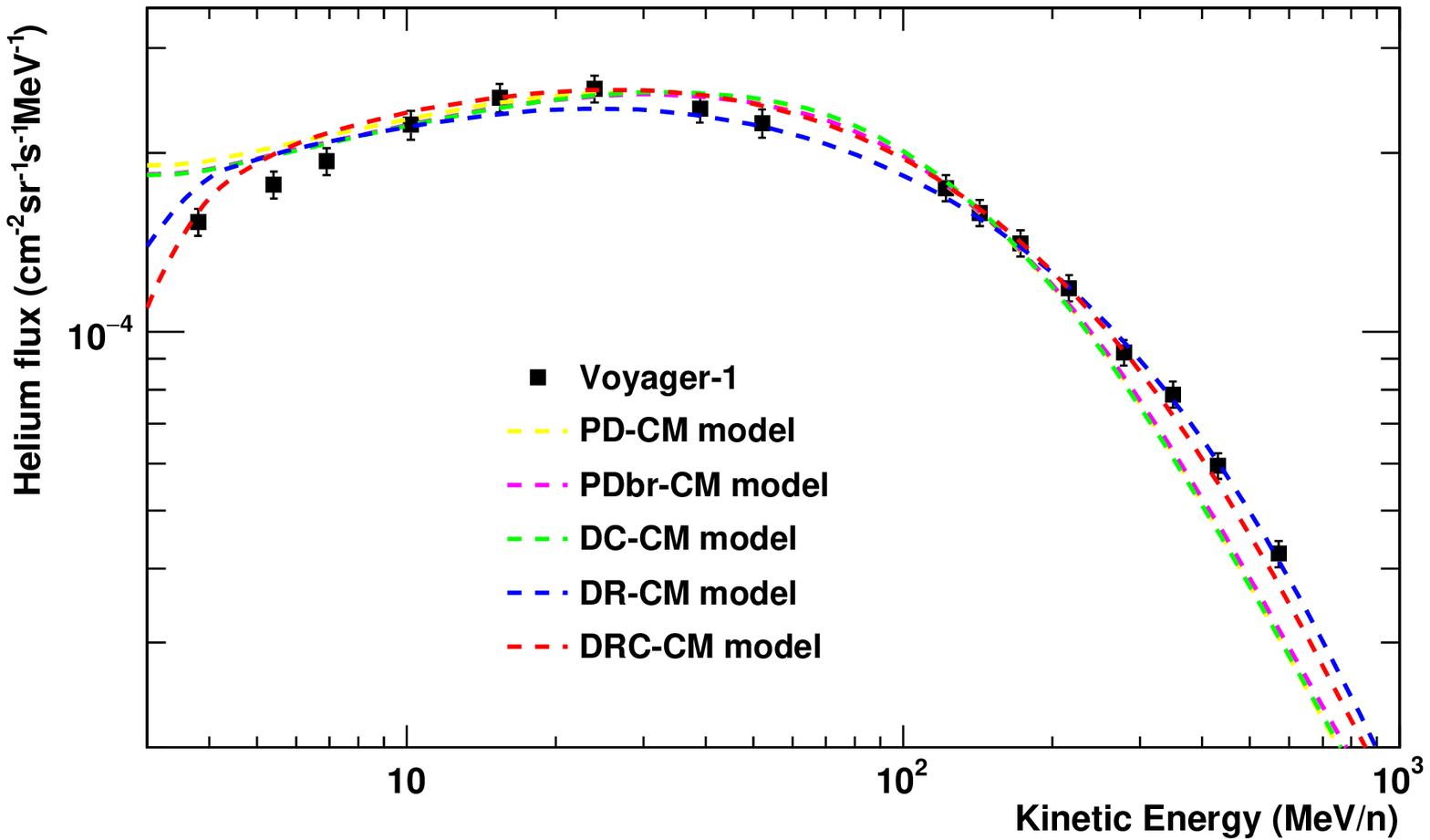}
\caption[The $^{2}\text{H}/^{4}\text{He}$ ratio and the helium flux compared with various models in the $\chi^{2}$ study]{\footnotesize The $^{2}\text{H}/^{4}\text{He}$ ratio and the helium flux for the best-fit parameters of PD-CM, PDbr-CM, DC-CM, DR-CM and DRC-CM models as listed in Table~\ref{tab:chi2}. Data points are the measurements from PAMELA and Voyager-1}\label{fig:chi2}
\end{figure}

For the CM scenario, the reduced $\chi^{2}$ is equal to 1.91 for the PD-CM model. This model does not agree with the $^2$H/$^4$He ratio measured by PAMELA and the unmodulated fluxes measured by Voyager-1. By introducing a break in the diffusion coefficient as in the PDbr-CM model, the reduced $\chi^{2}$ decreases to 1.55. Compared with the PD-CM model, the PDbr-CM model fits the interstellar proton flux better. However, as shown in FIG.~\ref{fig:chi2}, it still fails to reproduce the data of $^2$H/$^4$He ratio and interstellar helium flux. The reason the PDbr-CM model presents an acceptable overall $\chi^{2}$ is because that it overfits the PAMELA proton data. Therefore, though diffusion is the main process which CRs experience in Galaxy, it is not the sole CR propagation process. The DC-CM model also displays estrangements with the $^2$H/$^4$He ratio and the unmodulated helium flux. Moreover, both the DC-CM model and the PDbr-CM model yield rather high values of $\delta_{2} \sim 0.8$, which may conflict with the CR anisotropy~\cite{Strong2007}. Unlike the PDbr-CM and DC-CM models, the DR-CM and DRC-CM models can generally reproduce all the data. It implies that the reacceleration process, which is usually suggested to explain the behavior of the low energy B/C ratio, is also preferred by the $Z\leq2$ data. Compared with the DR-CM model, the DRC-CM model with one more parameter achieves a smaller $\chi^{2}$ value equal to 1.21, allowing a more satisfactory description of data.

\begin{table}[!htb]\small
\centering
\begin{tabular}{l  c | c }
\hline \hline
Parameters & Prior range & Prior type \\
\hline
$D_{0} $ ($10^{28}$cm$^{2}$\,s$^{-1}$) & [1.0, 10] & \multirow{13}{*}{Uniform}\\
$\delta$& [0.2, 0.6] &\\
$\eta$&[1.0,2.0]& \\
$v_{A} $ (km\,s$^{-1}$) & [20, 40] & \\
$\text{d}V/\text{d}z$ (km\,s$^{-1}$\,kpc$^{-1}$) & [10, 40] & \\
$\nu_{1}$ & [1.5, 2.1] & \\
$\nu_{2}$ & [2.1, 2.7] & \\
$\rho_{br}$~(GV) & [5.0, 10.0] & \\
$\Delta_{\text{He}}$& [-0.08,0] & \\
$X_{\text{He}}$& [0.06,0.08] & \\
$\phi_{0}$~(GV) & [0.1, 0.4] & \\
$\phi_{1}$~(GV) &[4, 12] & \\
\hline
\end{tabular}
\caption[Priors for parameters in the Bayesian study] {\footnotesize Priors on propagation parameters, source parameters and solar modulation parameters. The prior on each parameter is uniform to assign equal probabilities on all the possible values within the prior range.}\label{tab:priors}
\end{table}

The DR-CM model and the DRC-CM model are further studied in the Bayesian analysis. Priors are specified on the free parameters to restrict them in physically reasonable regions, as shown in Table~\ref{tab:priors}. The results are summarized in Table~\ref{tab:bayesian_results}, including the logarithm of Bayesian evidence for each model and the posterior mean with standard deviation on parameters. The differences in log-evidence between these two models is $\mathrm{ln} \left(Z_{2}/Z_{1}\right) \to 24$. An inclusion of convection significantly increases the evidence, suggesting that the DRC-CM model is the ``best'' one among all the models under study.

\begin{table}[!htb]\small
\centering
\begin{tabular}{l  c  c }
\hline \hline
Parameter & DR & DRC \\
\hline
$D_{0} $ ($10^{28}$cm$^{2}$\,s$^{-1}$) &  $8.25\pm0.20$ & $3.9\pm0.5$\\
$\delta$& $0.237\pm0.022$ & $0.29\pm0.04$\\
$\eta$& $1.50\pm0.06$ & $1.52\pm0.06$\\
$v_{A}$ (km\,s$^{-1}$)&$29.2\pm1.6$& $30.6\pm1.4$\\
$\text{d}V/\text{d}z$ (km)\,s$^{-1}$\,kpc$^{-1}$&--- & $31\pm4$\\
$\nu_{1}$ &  $1.695\pm0.019$ & $1.766\pm0.026$\\
$\nu_{2}$ & $2.531\pm0.021$ & $2.544\pm0.027$\\
$\rho_{br}$ (GV) & $6.8\pm0.4$ & $7.3\pm0.4$\\
$\Delta_{\text{He}}$ & $-0.060\pm0.005$ & $-0.046\pm0.005$\\
$X_{\text{He}}$& $0.0675\pm0.0012$ & $0.0704\pm0.0013$\\
$\phi_{0}$ (GV) & $0.265\pm0.013$ & $0.224\pm0.013$\\
$\phi_{1}$ (GV) & $8.1\pm0.7$ & $9.2\pm0.7$\\
log-evidence&$-191.91\pm0.23$&$-167.58\pm0.23$\\
\hline
\end{tabular}
\caption[The posterior mean with standard deviation for DR and DRC models under the GALPROP-CM framework in the Bayesian study]{\footnotesize The posterior mean with standard deviation of the free parameters for the DR-CM and DRC-CM models. The derived logarithmic evidences are also given.}
\label{tab:bayesian_results}
\end{table}

Parameter evaluations from the Bayesian method are consistent with the results obtained from the $\chi^{2}$ analysis. The ``best'' DRC-CM model yields a diffusion spectral index $\delta=0.29\pm0.04$, encouraging a Kolmogorov-type magnetic turbulence. We test a different fit using default GALPROP $^{2}$H and $^{3}$He production cross-sections. The value of $\delta$ can be upshifted by 0.07, but with much worse goodness of fit. Even in this case, it still favors a Kolmogorov-type turbulence. The \textit{Alfv$\acute{e}$n velocity} $v_{A}$ allowed for the DRC-CM model is around $30.6\pm1.4$~km/s. A strong reacceleration level is needed. In our analysis, $V_{0}$ is fixed at 0~km/s for convection models. This is not an arbitrary choice. If we free $V_{0}$ in the fit, it converges at zero. It hints that adopting a constant convection velocity may rest on an incorrect assumption. Nevertheless, in our work we fix $V_{0}$ to reduce the number of free parameters and diminish the calculation time. The estimated $\text{d}V/\text{d}z$ value is equal to $31\pm4$~km\,s$^{-1}$\,kpc$^{-1}$, suggesting a strong convection velocity. Constraints are also placed on the injection parameters. The results show that a softening on injection spectrum at $7.3\pm0.4$~GV is necessary for the DRC-CM model to recover the data. For all the models, without a break on the spectral index, the derived $\chi^{2}$ values are highly enlarged. This implies that a single power law is insufficient to reproduce the data. The posterior mean values of $\nu_{1}$ and $\nu_{2}$ are $1.766\pm0.026$ and $2.544\pm0.027$, respectively. A sum of $\delta+\nu_{2}$ is around 2.8 to hold the measured high energy proton flux.

\begin{figure}[!htb]
\includegraphics[width=8.5cm]{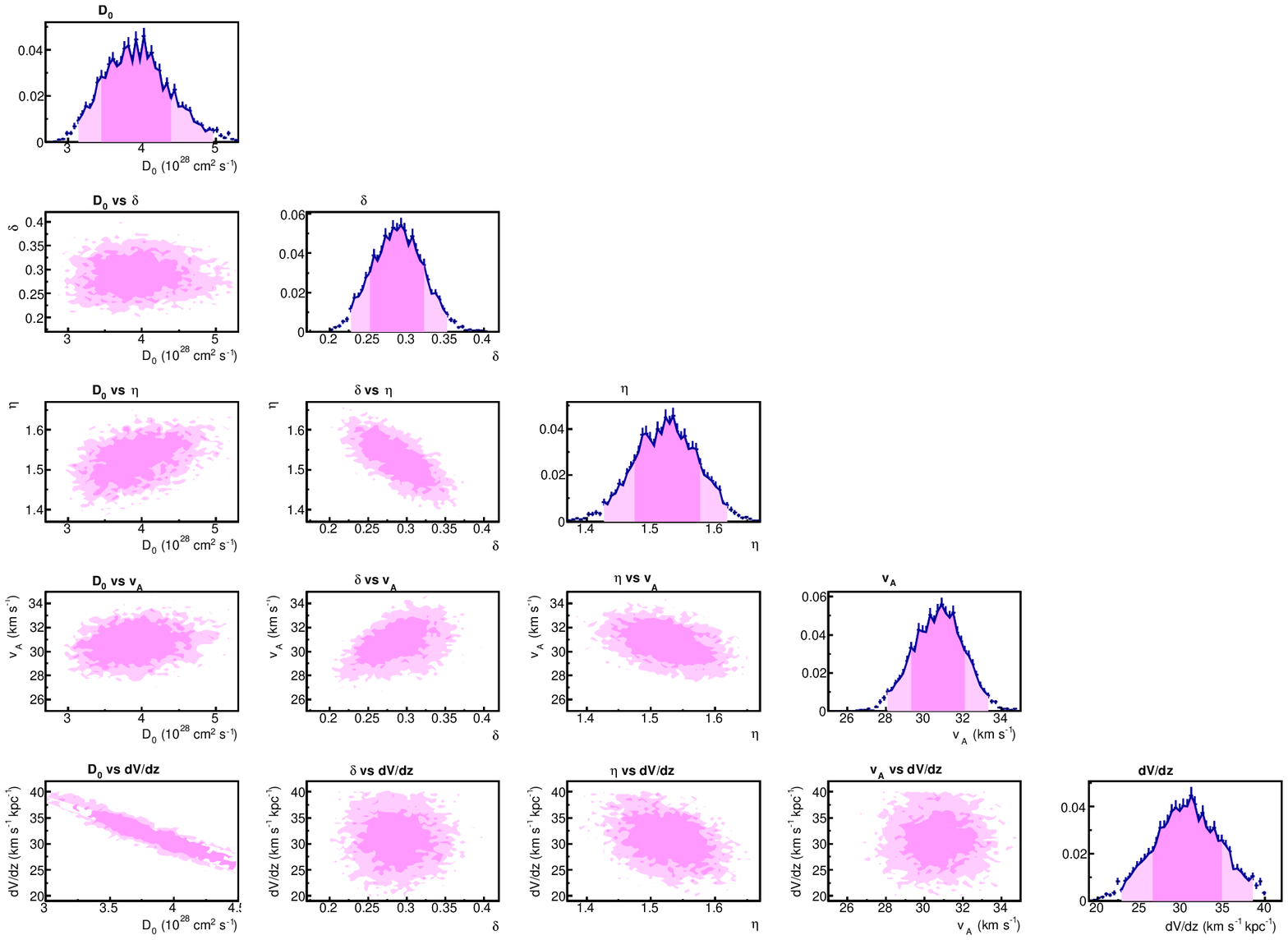}
\includegraphics[width=8.5cm]{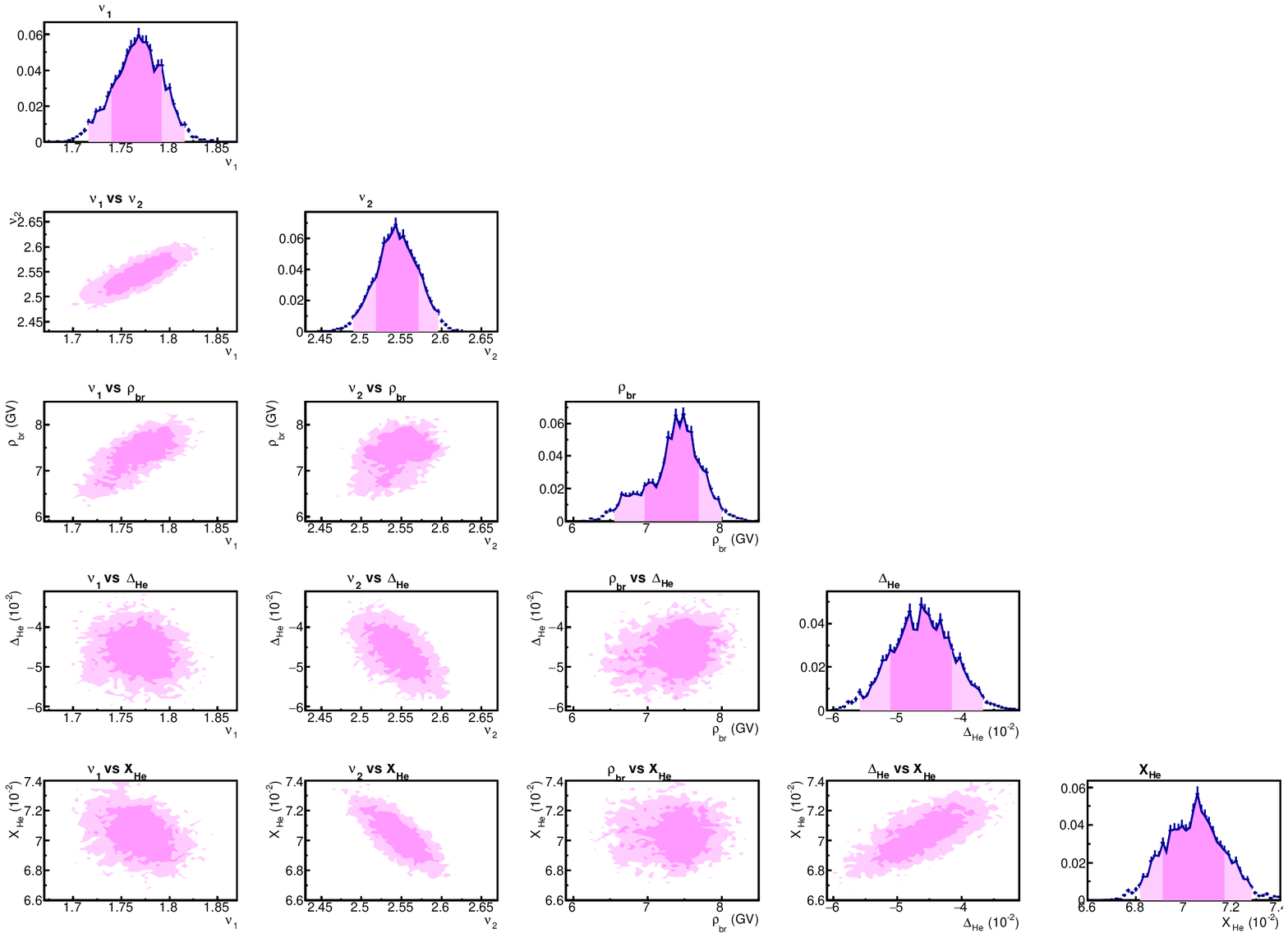}
\caption[The parameter pdfs for the DR model]{\footnotesize The 1D (diagonal) and 2D (off-diagonal) marginalized posterior p.d.f.s of the transport and source parameters for the DRC-CM model (as shown in Table~\ref{tab:bayesian_results}). The dark/light purple area represents the 68\%/95\% credible interval.}\label{fig:DRC_pdf_prog}
\end{figure}

The 1D and 2D marginalized posterior p.d.f.s of free propagation
and source parameters for the DRC-CM model are given in FIG.
\ref{fig:DRC_pdf_prog}. As shown, a negative correlation exists
between $\eta$ and $\delta$. This is expected from
Equation~\ref{eq:diff_coeffieient}. In order to reproduce the low
energy secondary-to-primary ratios, a larger $\eta$ requires a
smaller $\delta$. The convection velocity $\text{d}V/\text{d}z$ is
found to be negatively correlated with $D_{0}$. For CR particles,
a fast diffusion needs a slow convection to maintain a sufficient
transport duration in Galactic halo. The relationships between
other transport parameters do not present arresting correlations.
Considering the source parameters, a noticeable negative relation
appears between $\nu_{2}$ and $X_{\text{He}}$. This is because a
flatter injection spectrum will deviate more from the measured
helium flux if it adopts a lower value of helium abundance
normalization. The injection indices $\nu_{2}$ and $\nu_{1}$
exhibit a positive correlation. This can be understood by taking
into account their relations with $\delta$. To achieve the
observed propagated CR slope, a larger $\delta$ leads to a flatter
injection spectrum both at low rigidities and at high rigidities.
This causes a positive correlation between $\nu_{2}$ and
$\nu_{1}$. The degeneracy between $\delta$ and $\nu_{2}$ can be
broken by using high energy secondary-to-primary ratios. In this
work, to clarify the reliability of estimated $\delta$, we use the
$\bar{\text{p}}/\text{p}$ ratio to do a cross-check in
Sect.~\ref{sec:sp}.

\begin{figure*}[!ht]
\centering
\includegraphics[width=8.5cm]{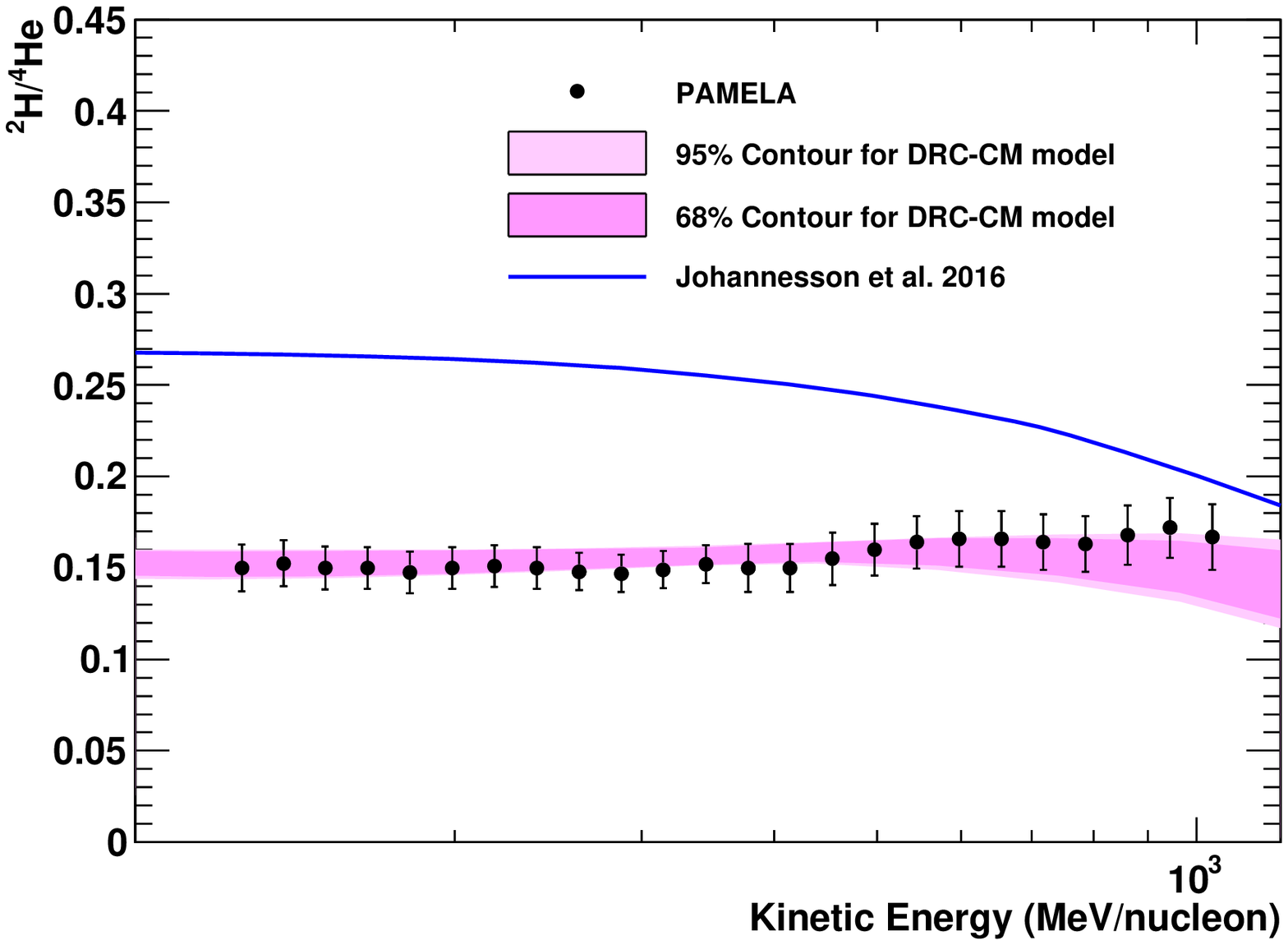}
\includegraphics[width=8.5cm]{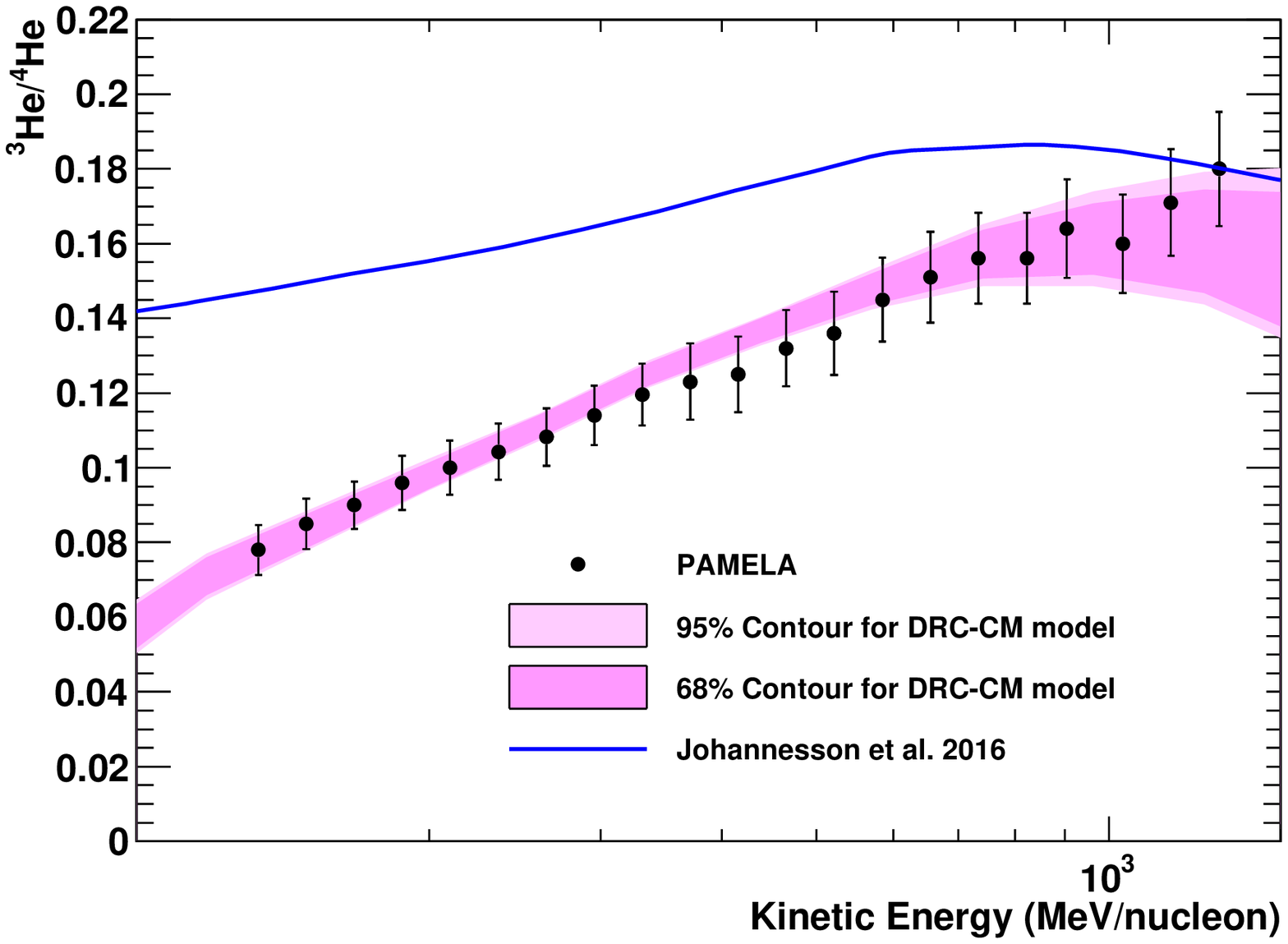}
\includegraphics[width=8.5cm]{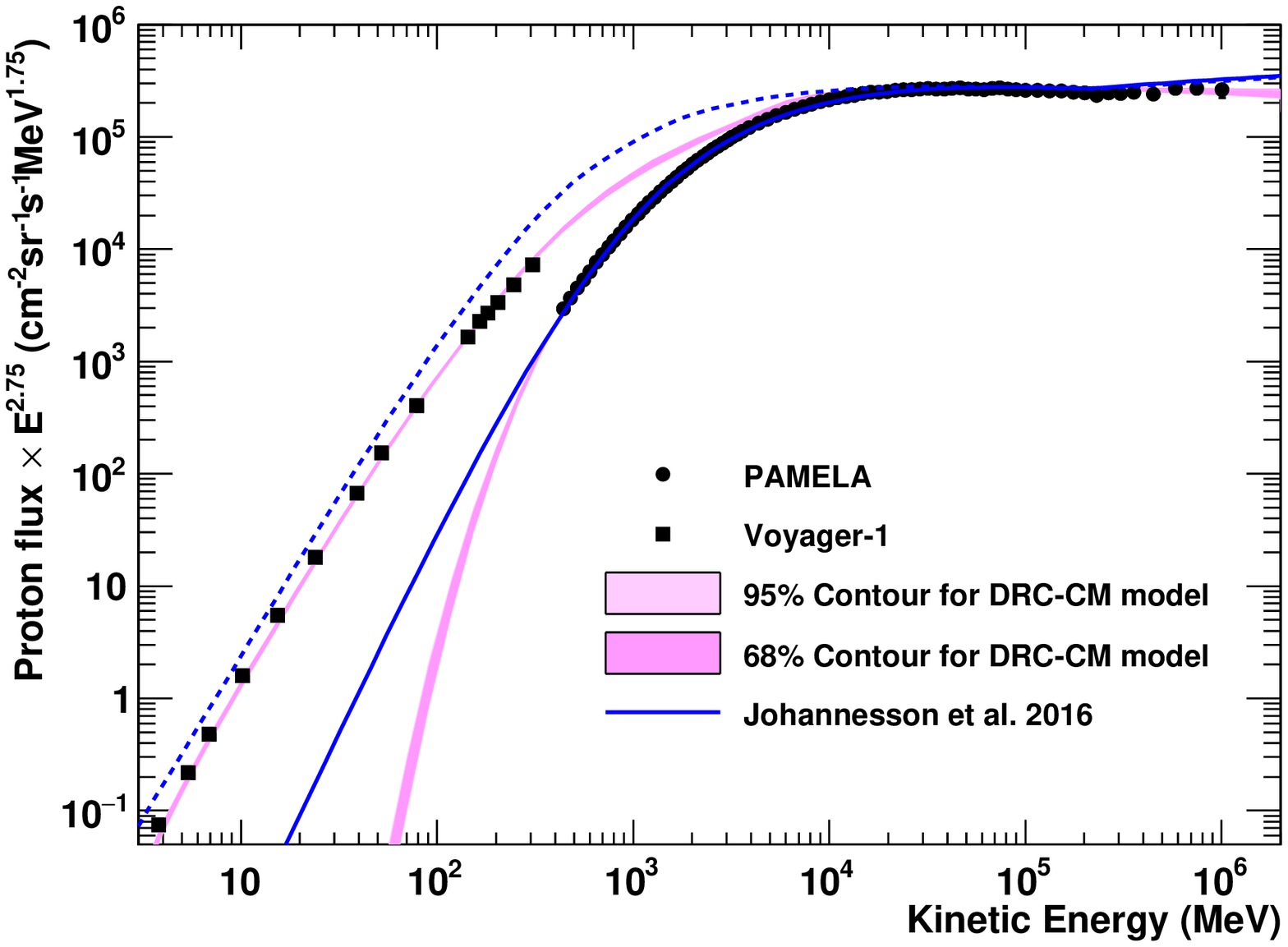}
\includegraphics[width=8.5cm]{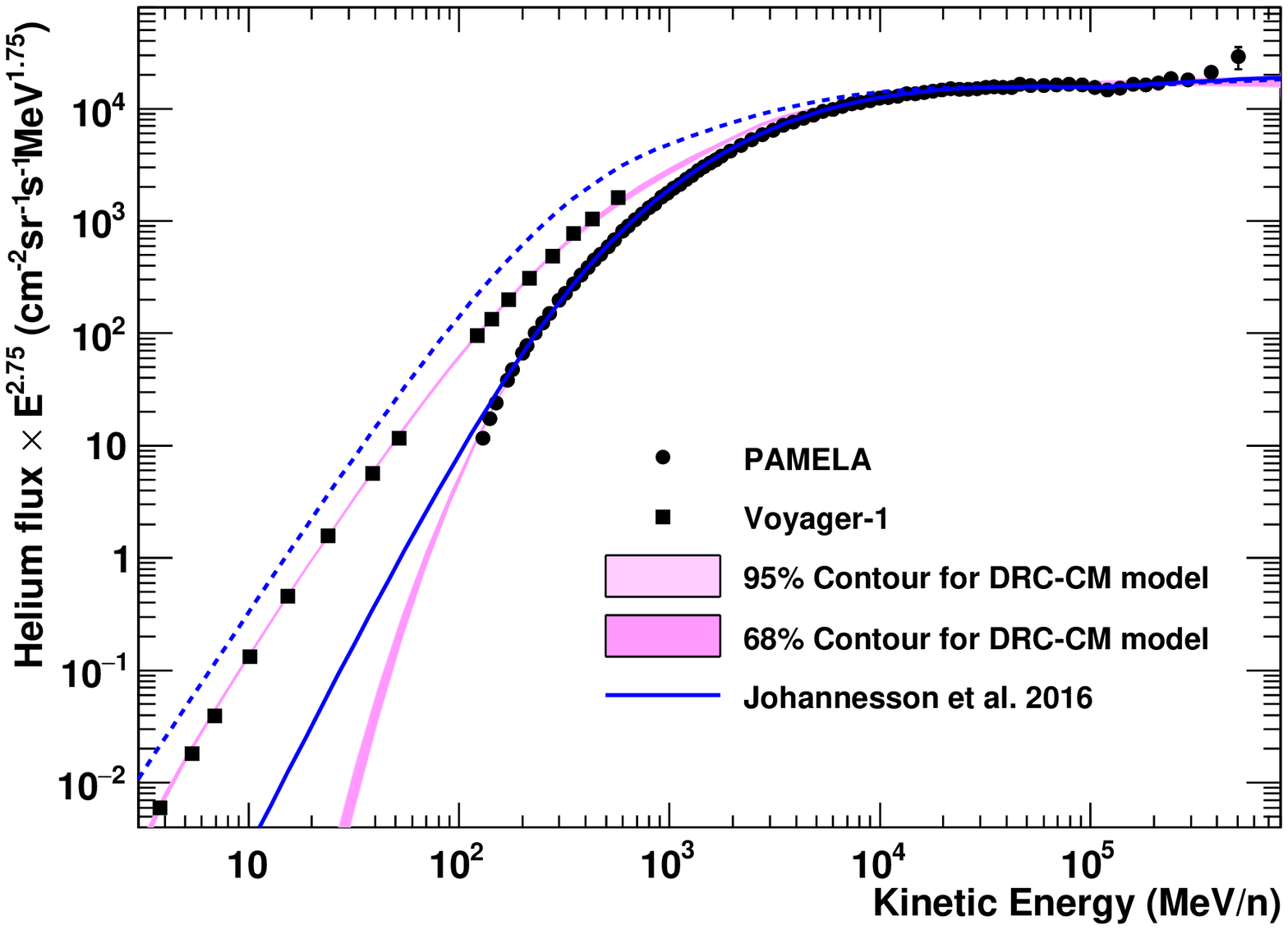}
\caption[The $^{2}\text{H}/^{4}\text{He}$ ratio, the $^{3}\text{He}/^{4}\text{He}$ ratio, the proton flux and the helium flux in the Bayesian study]{\footnotesize The $^{2}\text{H}/^{4}\text{He}$ ratio, the $^{3}\text{He}/^{4}\text{He}$ ratio, the proton flux and the helium flux for the best DRC-CM model as listed in Table~\ref{tab:bayesian_results}. The dark/light purple area represents the 68\%/95\% credible interval. The blue line is computed using the best-fit parameters derived in~\cite{Johannesson2016} by using only low-mass data (p, He, $\bar{\text{p}}$). In bottom panels, the upper purple band and the dashed line represent the calculated unmodulated spectra. }\label{fig:bayesian}
\end{figure*}

The theoretical calculations in 68\% and 95\% credible intervals for the fitted CR spectra are shown in FIG. \ref{fig:bayesian}. It can be found that the DRC-CM  model can satisfactorily recover the PAMELA data including the $^2$H/$^4$He ratio, the $^3$He/$^4$He ratio, the proton flux and the helium flux. Meanwhile, this model also accommodates the interstellar CR spectra released by Voyager-1. Our results are compared with a diffusive reacceleration model proposed in an earlier paper~\cite{Johannesson2016}. That model was claimed as the best-fit model for the low-mass CR elements. However, it shows dramatic disagreements with
the observed $^2$H/$^4$He, $^3$He/$^4$He ratios and the unmodulated fluxes. The bias on their parameters is due to the employment of an inappropriate description on solar modulation, as discussed in Sect. \ref{sec:chi2_FF}.

\subsection{Comparison with other secondary-to-primary ratios} \label{sec:sp}

\begin{figure}[!htbp]
\includegraphics[width=8.5cm]{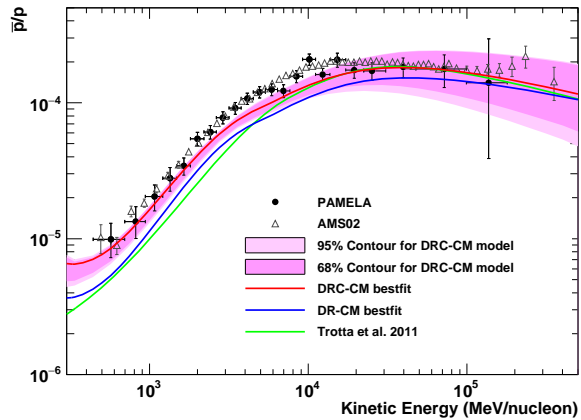}
\caption[The $\bar{\text{p}}/p$ in the Bayesian study]{\footnotesize The $\bar{\text{p}}/\text{p}$ ratio for the best DRC-CM model as listed in Table~\ref{tab:bayesian_results}. The dark/light purple area represents the 68\%/95\% credible interval for the DRC-CM model. The red line is derived by using the best-fit parameters for the DRC-CM model. The blue line presents the predictions for the DR-CM. The green line shows the theoretical calculation for the best-fit DR model derived in~\cite{Trotta2011}.}
\label{fig:bayesian_pbarp}
\end{figure}

The prediction for the $\bar{\text{p}}/\text{p}$ ratio is calculated. The default $\bar{\text{p}}$ production cross-sections embedded in GALPROP~\cite{Tan1983a, Tan1983b, Moskalenko2002} is used here. As shown in FIG. \ref{fig:bayesian_pbarp}, compared with the conventional DR model suggested in~\cite{Trotta2011}, which adopted a force-field approximation, the DRC-CM model (see Table~\ref{tab:chi2}) yields 30\%$\sim$60\% higher $\bar{\text{p}}$ production around a few GeV. It seems that an inclusion of a convection mechanism and incorporating a charge-dependent heliospheric modulation model both influence substantially to the $\bar{\text{p}}/\text{p}$ ratio. The DRC-CM model agrees fairly well with the PAMELA data below 1~GeV and above 10~GeV. At moderate energies, the DRC-CM model yields a slightly lower $\bar{\text{p}}/\text{p}$ ratio. This discrepancy might be caused by the inaccuracy of $\bar{\text{p}}$ production cross-sections~\cite{Johannesson2016, Lin2017}. Nevertheless, at energies larger than 20~GeV, the calculated $\bar{\text{p}}/\text{p}$ ratio for the DRC-CM is in an excellent agreement with the observations by both PAMELA and AMS-02~\cite{Aguilar2016a}. This is an indication that the derived value of $\delta$ remains robust for all the Z$\leq$2 particles.

\begin{figure}[!htbp]
\includegraphics[width=8.5cm]{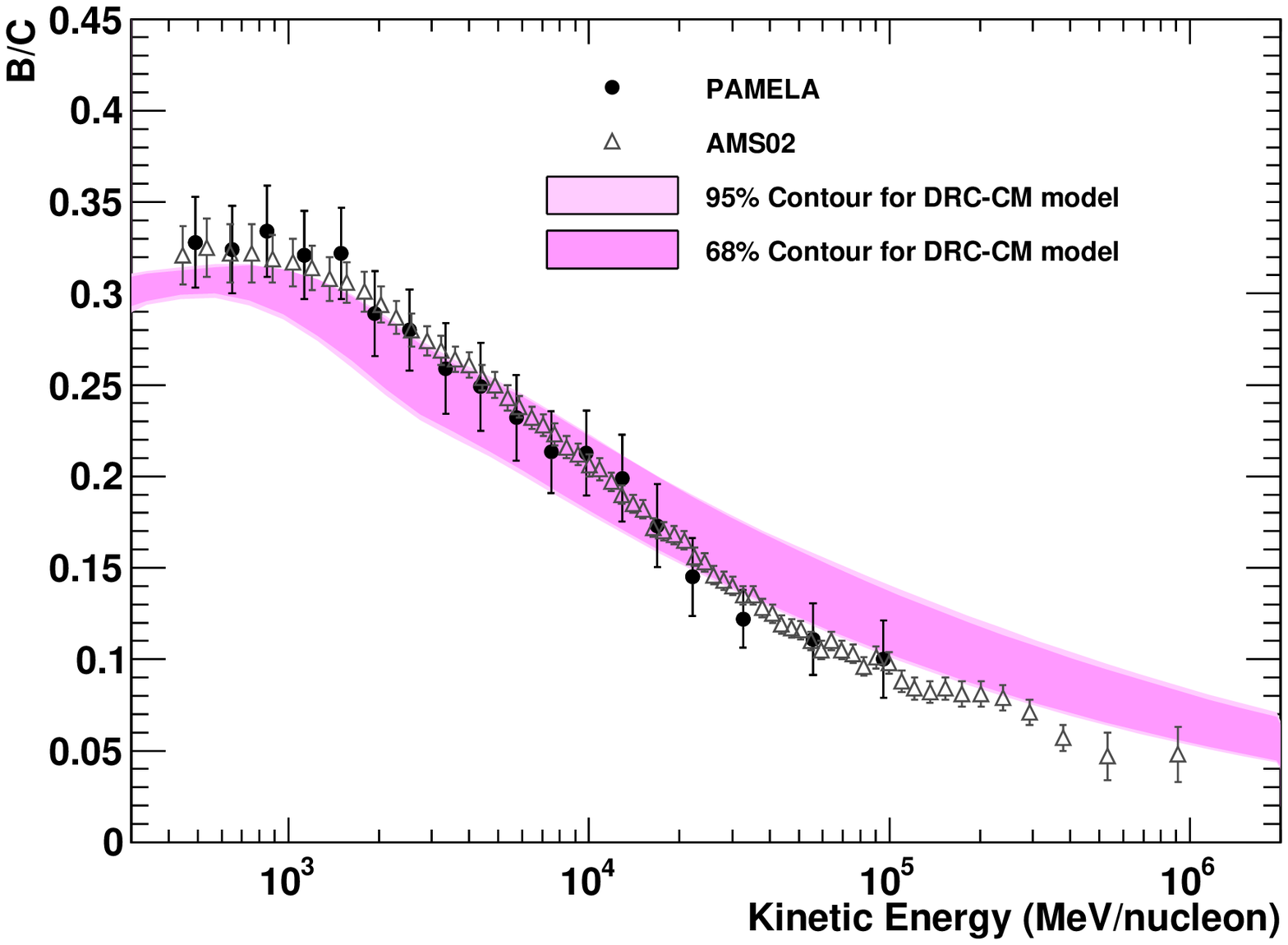}
\caption[The B/C in the Bayesian study]{\footnotesize The B/C ratio for the best DRC-CM model as listed in Table~\ref{tab:bayesian_results}. The dark/light purple area represents the 68\%/95\% credible interval for the DRC-CM model.}
\label{fig:bayesian_bc}
\end{figure}

The evaluated parameters for the DRC-CM model is also used to make a calculation of the B/C ratio. As shown in FIG.~\ref{fig:bayesian_bc}, the DRC-CM model can reproduce the PAMELA data~\cite{Adriani2014}. But the prediction is slightly higher than the AMS-02 data~\cite{Aguilar2016b} at high energy. The slope of the theoretical B/C ratio seems slightly flatter than the AMS-02 data. It may indicate a larger $\delta$ for heavy-mass elements. This discrepancy might also relate to several possible factors, for example, the systematic inconsistencies between different data-sets, the possible systematic uncertainties in production cross-sections and/or in solar modulation. These systematic effects and a joint analysis on both light and heavy elements will be further explored in our future work.

\subsection{Comparison with other analyses}

Our results agree fairly well with the estimates given in~\cite{Korsmeier2016}. By using the $\bar{\text{p}}/\text{p}$ ratio and the Z$\leq$2 primaries, they found $\delta=0.30\pm^{0.03}_{0.02}(stat)\pm^{0.10}_{0.04}(sys)$ and $v_{A}=25\pm2$~km/s. Compared with the values of $\delta$ obtained by using the B/C ratio, for example $\delta=0.395\pm0.025$ and $\sim0.35$ as presented in~\cite{Boschini2017} and~\cite{Cuoco2017b}, our estimate is slightly lower. This is also inferred from FIG.~\ref{fig:bayesian_bc}, as discussed in Section~\ref{sec:sp}. Taking into account the possible systematic uncertainties, light and heavier nuclei may still have compatible diffusion process. Concerning the convection process, a constant convection velocity was assumed in~\cite{Korsmeier2016}. And in~\cite{Boschini2017}, they found the convection velocity with $V_{0}=12.4\pm0.8$~km/s and $dV/dz=10.2\pm0.7$~km\,s$^{-1}$\,kpc$^{-1}$. In contrast, we find that $V_{0}$ converges at 0~km/s and a larger $dV/dz$ is required. In~\cite{Boschini2017}, we note that their constraints on transport parameters depended on the approximate solar modulation parameters. As a consequence, a different choice on modulation parameters might lead to a different estimation on CR propagation parameters. In our analysis, we free simultaneously the solar modulation parameters. The difference in the fitting procedure may be responsible for the inconsistencies on convection. Anyway, our result on $V_{0}$ is more appealing for the sake of the continuity on convection velocity at Galactic plane.

\section{Conclusion}

For the first time, we use the PAMELA measurements of hydrogen and
helium isotopes to perform an elaborate study and a comparison on
cosmic ray propagation models. Accurate and reliable constraints
are generated by using the $^2$H/$^4$He and $^3$He/$^4$He ratios
combined with the Z$\leq$2 primaries. Several hints can be
inferred from our results. First, a global fit on both the
modulate and unmodulated data disfavors the widely used
force-field approximation. Instead, the tension between CR
propagation models and the observations can be released by
incorporating the time-, charge-, rigidity-dependent CM solar
modulation model. Second, various models are comprehensively
studied and compared by using statistical methods. From a quick
$\chi^{2}$ minimization analysis, it is found that only models
including the reacceleration effect can accommodate individual
ratios and fluxes. A Bayesian analysis provides evidences which
strongly support the DRC-CM model over the DR-CM model. The
$\bar{\text{p}}/\text{p}$ ratio not involved in the fitting
procedure further demonstrates the plausibility of our DRC-CM
model and the robustness of the estimated diffusion spectral index
$\delta$. At least, the DRC-CM model show good agreements with all
the Z$\leq$2 data. Third, the value of $\delta$ is determined to
be close to 1/3, which indicates a Kolmogorov-type turbulence.
This estimated $\delta$ based on Z$\leq$2 data is slightly smaller
than that value derived from the B/C ratio. However, systematics
uncertainties may contribute to this discrepancy, which will be
further investigated in the future.

\section{Acknowledgements}

We thank Xiao-jun Bi, Ilias Cholis, Su-jie Lin, Antje Putze and Stefano Della Torre for useful discussions. This work is supported by the Joint Funds of the National Natural Science Foundation of China (Grant No. U1738130) and the National Natural Science Foundation of China (No. 11303023 and No. 11475149). The use of the high-performance computing platform of China University of Geosciences is gratefully acknowledged.

\end{document}